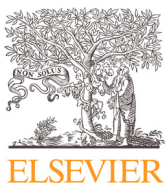
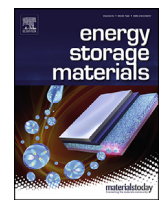

# Scalable spray-coated graphene-based electrodes for high-power electrochemical double-layer capacitors operating over a wide range of temperature

Mohammad Akbari Garakani [a,‡], Sebastiano Bellani [a,‡], Vittorio Pellegrini [a,b,*], Reinier Oropesa-Nuñez [b], Antonio Esau Del Rio Castillo [a], Sara Abouali [a], Leyla Najafi [a], Beatriz Martín-García [a], Alberto Ansaldo [a], Paolo Bondavalli [c], Cansunur Demirci [d], Valentino Romano [e], Elisa Mantero [a], Luigi Marasco [a], Mirko Prato [f], Gaetan Bracciale [c], Francesco Bonaccorso [a,b,*]

[a] *Graphene Labs, Istituto Italiano di Tecnologia, via Morego 30, 16163 Genova, Italy*
[b] *BeDimensional Spa, via Lungotorrente Secca 3d, 16163 Genova, Italy*
[c] *Chemistry and Multifunctional Materials Lab, Technology and measurements Department, Thales Research and Technology, 91767, Palaiseau, France*
[d] *NanoChemistry, Istituto Italiano di Tecnologia, via Morego 30, 16163 Genova, Italy*
[e] *Dipartimento di Scienze Matematiche ed Informatiche, Scienze Fisiche e Scienze della Terra, Università di Messina, Viale F. Stagno D'Alcontres 31, 98166 Messina, Italy*
[f] *Materials Characterization Facility, Istituto Italiano di Tecnologia, via Morego 30, 16163 Genova, Italy*

A B S T R A C T

Advancements in electrochemical double-layer capacitor (EDLC) technology require the concomitant use of novel efficient electrode materials and viable electrode manufacturing methods. Cost-effectiveness, scalability and sustainability are key-drivers for fulfilling product development chain accepted by worldwide legislations. Herein, we report a scalable and sprayable "green" electrode material-based ink based on activated carbon and single-/few-layer graphene (SLG/FLG) flakes. We show that, contrary to commercial reduced graphene oxide, defect-free and flat SLG/FLG flakes reduce the friction of ions over the electrode films, while spray coating deposition of our ink maximises the electrolyte accessibility to the electrode surface area. Sprayed SLG/FLG flakes-based EDLCs display superior rate capability performance (*e.g.*, specific energies of 31.5, 23.7 and 12.5 Wh kg$^{-1}$ at specific powers of 150, 7500 and 30000 W kg$^{-1}$, respectively) compared to both SLG/FLG flakes-free devices and commercial-like EDLCs produced by slurry-coating method. The use of SLG/FLG flakes enables our sprayed EDLCs to operate in a wide range of temperature (−40/+100°C) compatible with ionic liquid/organic solvent-based electrolytes, overcoming the specific power limits of AC-based EDLCs. A prototype EDLCs stack consisting of multiple large-area EDLCs, each one displaying a capacitance of 25 F, demonstrates the industrial potential of our technology.

## 1. Introduction

Electrochemical double-layer capacitors (EDLCs) are the class of supercapacitors (SCs) that exclusively rely on a reversible accumulation of charges at the interface of the electrodes and electrolyte through ion adsorption [1,2]. Thanks to the high specific surface area (SSA) of the EDLC electrodes (preferably > 500 m$^2$ g$^{-1}$), the EDLCs normally exhibit 10 to 100 times higher specific energy (typically ranging between 1–10 Wh kg$^{-1}$) compared to those of conventional electrolytic capacitors (typically ranging between 0.01-0.3 Wh kg$^{-1}$) [3,4]. The fast reversibility of ion adsorption processes allows the EDLCs to be charged and discharged in the timescales of second/minute [5]. Therefore, they can be effectively used in a variety of applications where high-specific powers (> 10 kW kg$^{-1}$) are required [6,7], such as energy regenerative braking systems [8,9], emergency power units in avionics [10] and trains [11,12], rapid energy storage/release [13] in electronics [14] and communication systems [13–15]. Moreover, EDLCs benefit from their exceptional cyclic life (~millions of cycles) [16,17] in comparison with batteries (a few thousands of cycles) [18–20]. In fact, rechargeable batteries mainly suffer from the degradation of the electrodes related to chemical reactions occurring during the charge and discharge processes. For instance, intercalation-alloying induced stresses can cause significant changes in the electrodes structure [21,22] and, subsequently, capacity fading [23]. Since the specific energy of commercial EDLCs (< 10 Wh kg$^{-1}$) is still significantly lower than that of the battery systems (*e.g.*, 30-70 Wh kg$^{-1}$ for Ni-Cd batteries [24], and 150-270 Wh kg$^{-1}$ for Li-ion batteries [23,25] and 50-120 Wh kg$^{-1}$ Ni–metal hydride batteries [24,26,27]), recent efforts have mainly focused on their energy enhancement. Obviously, superior specific energies have to be addressed while preserving the advantageous features of the EDLCs, in particular, their high rate capability (> 10 kW kg$^{-1}$) [1,2,6,7]. In this context, high-specific energy pseudo-capacitors (also named faradaic supercapacitors) have been proposed to shorten the energy gap between EDLCs and batteries. However, the electrochemical storage mechanism of the pseudo-capacitors relies on redox reactions, intercalation or electrosorption (surface redox reaction) processes [28,29], which limit both their cyclic stability and the maximum specific power relatively to the EDLCs.

In commercially available EDLCs, activated carbons (ACs) are typically used as electro-active materials because of their large SSA (between 500–3500 m$^2$ g$^{-1}$) as well as their low cost (<10 USD kg$^{-1}$) [30]. However, the presence of pores with diameter inferior to 2 nm (*i.e.*, micropores) [2,31] and "blocked micropores" (*i.e.*, inaccessible isolated micropores) [32–34] in the AC electrodes limits the electrolyte ions accessibility (only ∼1/3 of the SSA is effectively accessible by the electrolyte ions) [35–37], adversely affecting the EDLC rate capability (*i.e.*, the maximum operating power) [38]. These effects are further pronounced at low-temperature (< 0 °C) EDLC operation, since the increase of viscosity of the electrolytic media, with decreasing the temperature, drastically reduces the ion mobility in microporous network. Moreover, the electrodes based on AC are often made by coating an electrode material slurry, consisting of AC together with conductive agents and binders, on a current collector (*e.g.*, a metallic substrate) [39]. This deposition method may be unconvenient to finely control the build-up of high mass loading films with over 5 mg cm$^{-2}$, acceptable mechanical integrity and optimal specific electrochemical performance, especially when exploiting viable "green" solvents (including water) [40]. To circumvent these drawbacks, major strategies mainly pursue three different objectives: 1) to tailor the structures of the electrodes [3,41–43] towards balanced microporosity (pore size from 0.2 to 2 nm) and/or mesoporosity (pore size from 2 to 50 nm); [36,40,44] 2) to control the interaction between electrode materials, solvent molecules and ions; [28,29] and 3) to enlarge the operating voltage window of the electrolytes [30,31]. In this scenario, other nanostructured allotropes of carbon beyond ACs have been investigated as advanced EDLC active materials [45,46]. Among them, graphene represents an ideal active material [47–51] due to its high theoretical SSA (∼2630 m$^2$ g$^{-1}$, corresponding to a theoretical specific (gravimetric) capacitance ($C_g$) superior to 500 F g$^{-1}$ in both ionic liquid and aqueous electrolytes) [52] and its exceptional charge carrier mobility (∼10$^5$ cm$^2$ V$^{-1}$ s$^{-1}$) [53–55]. *De facto*, graphene flakes can also act as a conductive additive in AC-based electrodes [56–58], whereas carbon nanoparticles (including AC) can electrically bridge the graphene flakes, synergistically increasing the overall electrical conductivity of the electrodes. Moreover, it has been recently demonstrated that graphene can act as a friction-free "ion slide" when organic electrolytes are used [38], facilitating the ion transport into nano-porous AC networks, whose morphological properties still play the main role in determining the final EDLC performance. Therefore, the exploitation of graphene can prospectively accomplish the crucial challenge of the current EDLC technology, namely the ionic charge accumulation in pores smaller than the size of the electrolyte ions [33], as predicted by atomistic models of electrosorbed ions in nano-confined environments [35,59].

The impact of graphene-based materials in EDLCs has been limited so far by different factors. First, graphene flakes naturally tend to re-stack during their deposition process due to van der Waals forces [60–64]. Additionally, graphene flakes tend to orient parallel to the current collectors [64–66], resulting in high-extended electrolyte pathways ineffective for fast charge/discharge processes in prototypical vertical EDLC configurations (especially in those cases with high mass loading electrode materials) [38,67]. These issues limit the practical implementation of graphene in commercial EDLCs, in which desirable energy/power densities are achieved with a typical active material mass loading of ∼10 mg cm$^{-2}$, meaning that the mass of the active material has to be at least 30% of the total device weight [68]. In principle, the fabrication of graphene-based highly porous structures (*e.g.*, three-dimensional (3D) graphene foams [69–71], and graphene-based hydro/aerogels [72–74]) can mitigate the aforementioned drawbacks. However, such strategy typically involves template-assisted growths [75], and/or complicated and tedious chemical functionalization [76,77]. These aspects arise severe cost and scalability issues, constricting their direct implementation in high-throughput electrode manufacturing [78]. In this context, it is worth pointing out that the entire EDLC production chain has to address the worldwide legislations/regulations to reduce the human health/environmental impact [79]. Currently, EDLC production is based on hard-to-dispose F-containing binders (*e.g.*, poly(vinylidene difluoride) –PVDF– and polytetrafluoroethylene –PTFE–) and often uses hazardous, teratogen, irritating and/or toxic solvents/dispersants (*e.g.*, N-Methyl-2-pyrrolidone –NMP–). However, several countries, including USA and European Union (EU), have already limited the NMP usage to a minimum (see Annex XVII to Reach –Regulation concerning the Registration, Evaluation, Authorization and Restriction of Chemicals–), indicating that NMP shall not be used in mixtures in a concentration equal to or greater than 0.3%, unless the exposure of workers is ensured to be below the Derived No-Effect Levels –DNELs– of 14.4 mg m$^{-3}$ for exposure by inhalation and 4.8 mg kg$^{-1}$ per day for dermal exposure [80]. Interestingly, major producers of EDLCs have recently developed a solvent-free electrode production processes (*i.e.*, the so-called "*dry electrode technology*", which enable them to operate at high throughput using a minimal manufacturing footprint [81]. Prospectively, and as alternative to the above dry electrode technology, EDLC manufacturing based on "*green*" solvents (*i.e.*, environmentally friendly and unhazardous solvents) could also eliminate the costs related to the use and recycling of the organic solvents and related binders [82]. For example, NMP is more expensive than water (>2 US$/kg *vs.* <0.02 US$/L) (<0.02 US$/L), the cost of PVDF (8-10 US$/kg) is superior to both alcohol and water/alcohol soluble binder such as carboxymethyl cellulose (CMC) (< 5 US$/kg) [83,84] natural cellulose –NC– (< 2 US$/kg) and alginate (∼8 US$/kg) [79]. However, it is challenging to formulate EDLC electrode material pastes/inks with "*green*" solvents. In fact, the apolarity of graphitic surfaces determine the precipitation and/or aggregation of the active materials in the most viable "green" solvent (*e.g.*, water and alcohols), which result in instable inks/pastes and low-quality electrode films, respectively [85,86].

In order to overcome the aforementioned issues, we propose a sprayable "*green*" ink of EDLC electrode materials based on a mixture of AC and single/few-layer graphene (SFG/FLG) flakes as active materials. The sprayable ink is formulated in an environmentally friendly and low-boiling point (≤ 100 °C) solvent mixture (*i.e.*, ethanol/water –EtOH/H$_2$O– (50:50 vol/vol)), using carboxymethyl cellulose (CMC) as a water-soluble binder. By acting as dispersing agent for SLG/FLG flakes in EtOH/H$_2$O mixture, CMC effectively stabilizes the as-produced ink. Our ink formulation in EtOH/H$_2$O mixture also eliminates the high-temperature processes (> 100 °C) for solvent removal after its deposition. Additionally, wet-jet milling (WJM) exfoliation of graphite is used for a scalable production of SLG/FLG flakes (production rate of ∼0.4 g min$^{-1}$ on a single WJM apparatus) [87]. Differently from other chemical exfoliation methods [88–90], the WJM approach is effective to preserve the chemical quality, and thus the electrical and tribological properties, of the pristine graphite, since it avoids the formation of defects and functional groups in the as-produced SLF/FLG flakes [87]. Both the material formulation and the spray coating deposition of the ink are compatible with high-throughput roll-to-roll (R2R) coating process implemented on flexible substrates (including plastics, papers and textiles). Moreover, lateral force microscopy (LFM) measurements reveal that the flat morphology and long-range two-dimensional (2D) sp$^2$ order of the added SLG/FLG flakes can effectively facilitate the "sliding" of ions, facilitating the electrolyte transport. The careful exploitation of the tribological properties of graphene and other carbonaceous nanomaterials could offer a strategy to further improve the EDLC performances, which are still mainly determined by the AC network properties. Consequently, symmetric EDLCs made of spray-coated graphene-based electrodes show a superior rate capability compared to the graphene-free devices. The replacement of the SLG/FLG flakes with commercial reduced graphene oxide (RGO) in the ink composition causes a deterioration of the rate capability attained by the sprayed EDLCs. This result indicates that the defect-free structure and the flat morphology of the SLG/FLG flakes are crucial to provide suitable electrode properties for the realization of high-power EDLCs. In addition, spray coating deposition of the electrode material is effective to improve both specific energy and power of EDLC compared to those of commercial-like SCs produced by slurry-coating methods, as a consequence of the superior SSA shown by the sprayed electrode films. The chemical, electrical and tribological properties of our SLG/FLG flakes, as well as the spray-coating-based electrode fabrication, enable the EDLCs to efficiently operate in a wide range of temperature (−40/+100 °C) compatible with ionic liquid/organic solvent mixture-based electrolytes, overcoming the specific power limits of graphene-free devices. These results highlight a new industrially compatible use of pristine graphene in EDLCs operating efficiently at high specific powers and extended temperature ranges.

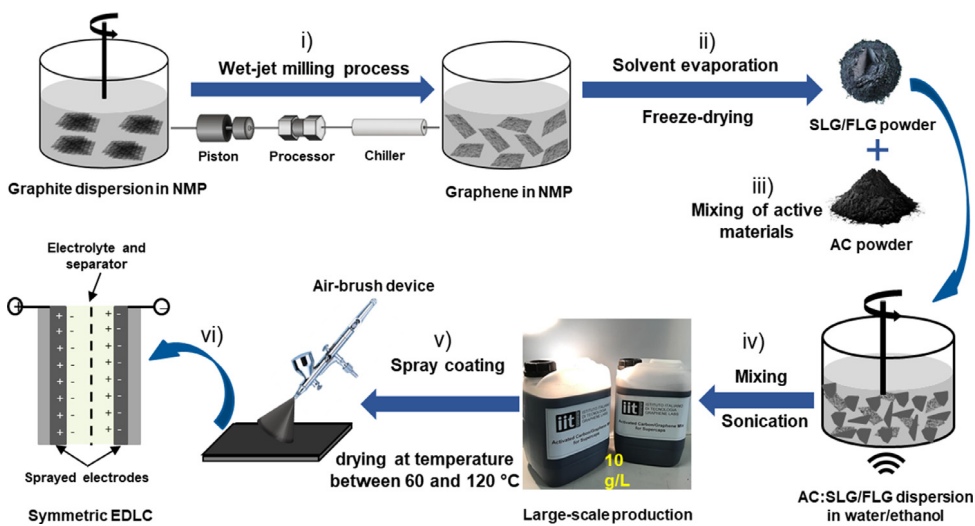

**Fig. 1.** Production chain of graphene-based EDLCs: i) WJM exfoliation of the graphite to produce SLG/FLG dispersion; ii) solvent evaporation and freeze-drying of the WJM-produced dispersion to obtain the SLG/FLG powder; iii) mixing of the active materials, *i.e.*, SLG/FLG and AC powders; iv) ultrasonication-assisted dispersal of the electrode materials in EtOH/$H_2O$ (50:50 vol/vol) mixture and addition of Na-CMC binder to formulate the electrode material ink; v) low-temperature (between 60 and 120 °C)-assisted spray coating of the electrode material ink to produce EDLC electrodes; vi) electrode assembly into symmetric EDLCs (coin-cell- or pouch-cell-type configurations).

## 2. Results and discussion

### 2.1. Production chain of graphene-based EDLCs

Fig. 1a schematically illustrates our EDLC production chain, including i) the WJM exfoliation of graphite to produce a SLG/FLG dispersion, ii) the solvent evaporation and freeze-drying of the WJM-produced dispersion to obtain the SLG/FLG powder, iii) the mixing of the active materials (*i.e.*, SLG/FLG and AC powders), iv) the ultrasonication-assisted dispersion of the electrode materials in EtOH/$H_2O$ (50:50 vol/vol) mixture and the addition of Na-CMC binder to formulate the electrode material ink; v) the low-temperature (between 60 and 120 °C)-assisted spray coating of the electrode material ink to produce EDLC electrodes; vi) the electrode assembly into EDLCs (coin-cell or pouch-cell configurations). To target industrial requirements, a large amount of SLG/FLG dispersion was produced by exfoliation of graphite through a scalable WJM method [38,87,91,92]. In particular, the time and the solvent volume required to produce 1 g of SLG/FLG ($t_{1\,gram}$) are 2.55 min and 0.1 L, respectively. The yield of exfoliation ($Y$) in this process, which is the weight ratio of the final graphitic flakes over the one of the initial graphite, is ~100% [87]. Such metric values are the best among those of the most common large-scale liquid-phase exfoliation (LPE) methods reported so far [93–100]. The details of the WJM process, as well as the SLG/FLG flakes characterization, are reported in the Supporting Information. Specifically, as shown by the atomic force microscopy (AFM) and transmission electron microscopy (TEM) analyses (Figure S2), both thickness and lateral size of the exfoliated flakes follow a lognormal distribution peaked at ~1.6 nm and ~460 nm, respectively [101]. Raman spectroscopy analysis (Figure S3) also revealed the single-/few-layer nature of the flakes, which show a long-range 2D-order of the $sp^2$ lattice. These results confirm that no in-plane defects have been induced into the graphene flakes during the exfoliation process, and defects are located only at the edges of the flakes [38,87,91]. In the X-ray photoelectron spectroscopy (XPS) C 1s spectrum of the SLG/FLG flakes (Figure S4), the dominating peak at 284.4 eV is attributed to $sp^2$ structures. This discards the formation of graphene oxide (GO), whose functional groups can affect the solvent and ion nanotribology on pristine graphene [102,103], as previously reported for high-power operating EDLCs [38]. The SLG/FLG powder, as obtained by freeze-drying method (see details in Supporting Information), was mixed with AC powder. Various material weight ratios were investigated for the AC:SLG/FLG mixtures (*i.e.*, 90:10 and 50:50) in order to optimise the ink formulation in terms of electrochemical performance. The active material binary mixture was dispersed *via* ultrasonication in EtOH/$H_2O$– (50:50 vol/vol) binary solvent (10 g L$^{-1}$ concentration), and Na-CMC was added as both dispersing agent and electrode binder with a 5 wt% of the total solid content. The as-formulated inks were deposited onto C-coated Al current collectors by a manual dual-action gravity feed airbrush or a custom-built automatized spray coater (see further details in Supporting Information, Table S1; electrodes hereafter named *sprayed*-AC:SLG/FLG). By adjusting the amount of the sprayed electrode material ink, electrode material mass loadings up to ~5 mg cm$^{-2}$ were deposited onto the current collectors. Noteworthy, these mass loadings are comparable to those adopted in commercial-like EDLCs, so that the device energy density values are not negatively affected (*e.g.*, they can be significantly decreased by a factor of over 10 for the mass loading of 1 mg cm$^{-2}$) [68]. Sprayed electrodes based on solely AC or SLG/FLG as active materials (named *sprayed*-AC and *sprayed*-SLG/FLG, respectively) and films obtained through conventional slurry casting deposition (*i.e.*, doctor blading) of the electrode materials, adopting conventional F-containing binders (*i.e.*, PVDF) and NMP (electrodes named *casted*-AC:SLG/FLG), were also produced as references. Finally, both coin- and pouch-cells were used to assemble two identical sprayed electrodes in symmetric EDLCs (hereafter named as their electrodes, *i.e., sprayed*-AC, *sprayed*-SLG/FLG, *sprayed*-AC:SLG/FLG, *casted*-AC:SLG/FLG).

### 2.2. Morphological characterization of the electrodes

The microstructure and the morphological features of the as-produced electrodes were assessed by scanning electron microscopy

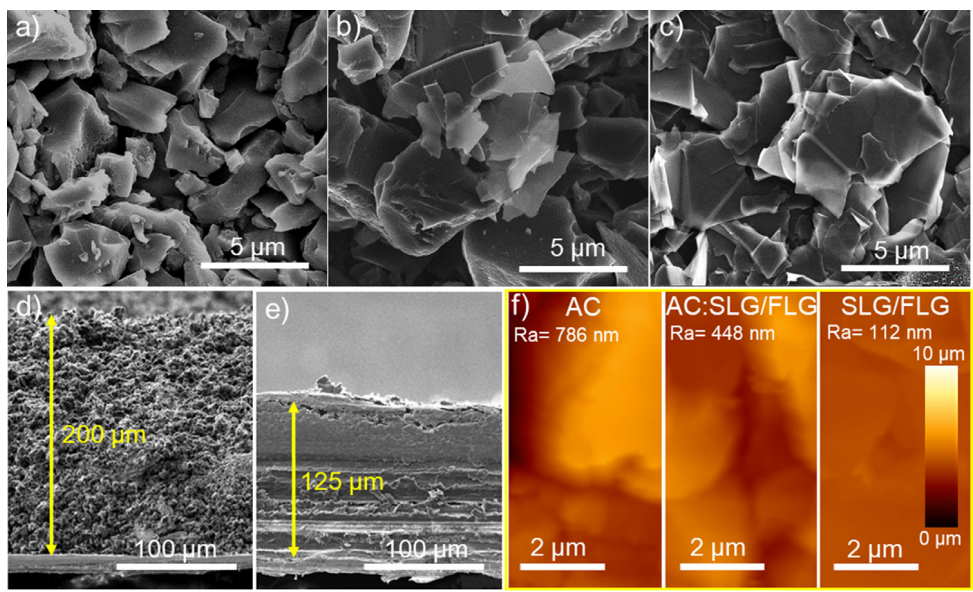

**Fig. 2.** SEM top-view images of the a) *sprayed*-AC, b) *sprayed*-AC:SLG/FLG (active material weight ratio of 90:10) and c) *sprayed*-SLG/FLG and SEM cross-sectional images of the d) *sprayed*-AC:SLG/FLG and e) *sprayed*-SLG/FLG. f) AFM results of the mentioned electrodes respectively, including the measured average roughness (Ra) for each electrode.

(SEM) measurements (Figure 2). The reference *sprayed*-AC shows a macro/mesoporous network made of AC particles with lateral dimensions from sub-micrometer to tens of micrometers (Fig. 2a), consistent with the material datasheets provided by the supplier [104]. The *sprayed*-AC:SLG/FLG (active material weight ratio of 90:10) resembles almost the same structure of the AC-based film due to the homogeneous distribution of the SLG/FLG flakes within AC particles (Fig. 2b). Differently, in the *sprayed*-SLG/FLG, the flakes are preferentially oriented horizontally to the substrate with a laminar structure (Fig. 2c), in agreement with previous reports on SLG/FLG films [58,62,63,105,106]. Notably, the homogeneity of the sprayed films is the results of stable, crack-free and aggregate-free electrode material films, determined by the dispersing role of the Na-CMC (subsequently acting as material binder). The electrode material of the ink progressively precipitates in absence of Na-CMC, showing sediment visible by eye after few minutes. We remark that Na-CMC is a linear polymer formed by the units of glucose linked *via* glycosidic $\beta$-(1 → 4) bonds [107,108]. The carboxymethylation of the cellulose backbone occurs when the latter reacts with an alkali solution (*e.g.,* NaOH) and chloroacetic acid or its Na-salt. Hydrophobic interactions between CMC backbones and the graphitic nanomaterials are formed when Na-CMC is added to the aqueous dispersion. The carboxylic groups of Na-CMC adsorbed on the graphite surface, stabilizing the graphene-based inks [106,109]. Fig. 2d, shows the cross-sectional SEM image of the *sprayed*-AC:SLG/FLG, evidencing a crack-free film with a thickness of ∼200 µm. The cross-sectional SEM image of the *sprayed*- SLG/FLG (Fig. 2e) displays a compact structure (film thickness ∼125 µm), as expected by the restacking of the SLG/FLG flakes. The roughness of the electrode was evaluated by AFM measurements (Fig. 2f). The average roughness (Ra) is 786, 448, and 112 nm for *sprayed*-AC, *sprayed*-SLG/FLG and *sprayed*-AC:SLG/FLG, respectively. These Ra values confirm the planarization induced by a preferential displacement of SLG/FLG flakes parallel to the substrate, resulting in a laminar structure agreeing with prior reports [64–66]. As previously demonstrated [38,67], such laminar-like compact arrangement of the SLG/FLG flakes is undesired in vertical-like EDLC configuration since it impedes an effective electrolyte access to the entire electrode film. Therefore, our morphological characterizations predict that the use of hybrid AC:SLG/FLG active materials is more effective than the use of the sole SLG/FLG.

The apparent SSA of the electrodes was evaluated by Brunauer-Emmet-Teller (BET) analysis [110] of physisorption measurements using Kr at 87 K (Figure S5) [111–114]. Since the quadrupole moment of $N_2$ ($0.27e \times 10^{-16}$ esu) [115] confines the access of the gas to micropores [111,113,114], Kr was preferred to $N_2$ in this experiment. More in particular, Kr suits for physisorption characterization of microporous carbon films [114,116,117], because it has not quadrupole moment and can access into the micropores with areas as small as its cross-sectional area (from 0.11 to 0.22 $nm^2$/atom at 77 K has been reported, depending on possible arrangements of the adsorbed species and the ranges of relative pressure) [116,117]. Moreover, Kr has a saturation pressure at 87 K lower than both $N_2$ and Ar [118], and this enables the detection of "small" pressure changes during adsorption processes occurring in micropores [118]. The calculated apparent BET specific surface area ($SSA_{BET}$) values are ∼2600, ∼2550 and ∼10 $m^2$ $g^{-1}$ for *sprayed*-AC, *sprayed*-AC:SLG/FLG and *sprayed*-SLG/FLG, respectively. The apparent $SSA_{BET}$ for AC electrode is higher than those provided by the material supplier (1666 ± 100 $m^2$ $g^{-1}$) [104], probably due to the different carrier gases used for the measurements, as well as the narrow nature of the pores (where the difficulty in the differentiation between monolayer/multilayer adsorption and micropore filling needs to be considered). The low $SSA_{BET}$ value measured for *sprayed*-SLG/FLG is due to the restacking of the flakes during their deposition [60–62]. In *sprayed*-AC:SLG/FLG, AC can act as effective spacers to avoid the restacking of the SLG/FLG flakes, as previously demonstrated in casted films [38]. The formation of high-packed wrinkled graphene-incorporated AC structures allows the most of the native SSA of graphene flakes to be fully used [119–121], resulting in film $SSA_{BET}$ comparable to the one of the AC films. Importantly, both the $SSA_{BET}$ and the pore size estimated for sprayed electrodes are superior to those of casted electrodes (*e.g.,* ∼1500 and ∼2100 for *casted*-AC and *casted*-AC:SLG/FLG), indicating that spray-coating deposition of active materials could be advantageous compared to the traditional slurry-coating method to achieve high SSA with optimal micro/mesoporosity balance. Non Localized Density Functional Theory (NLDFT) was used to calculate the pore size distribution of the electrodes measured by Kr physisorption at 87 K (Figure S6) [122,123]. The results show that *sprayed*-AC and *sprayed*-AC:SLG/FLG have similar nanoporous structures with pore sizes mainly smaller than 2 nm. The differential pore volume of *sprayed*-SLG/FLG is lower than those of the *sprayed*-AC and the *sprayed*-AC:SLG/FLG, suggesting the presence of "closed" micropores caused by the restacking of the SLG/FLG flakes, in agreement with BET analysis.

The electrical properties of the various sprayed electrodes were evaluated by measuring their sheet resistance ($R_{sheet}$) through four-probe method [124]. The obtained $R_{sheet}$ are ∼80, ∼70 and ∼40 $\Omega$ $sq^{-1}$ for *sprayed*-AC, *sprayed*-SLG/FLG and *sprayed*-AC:SLG/FLG, respectively. These results indicate that hybridised AC and SLG/FLG flakes synergistically decrease the $R_{sheet}$ of the electrodes. In fact, SLG/FLG flakes can act as electrically conductive additives [56–58], which are electrically bridged by AC [38]. Actually, the combination of carbon nanomaterials with different morphologies (0D, 1D and 2D) is an effective strategy for designing commercial conductive carbon-based pastes [63,125,126].

Beyond the electrical properties, nanoscale interactions of active materials with solvent and electrolytes may significantly affect the ultimate electrochemical performance of the EDLCs. Recent investigations [38] have shown that the combined control of the nanotribological properties of the active material/electrolyte system can propose advanced solutions to satisfy both energy and power density requirements in EDLCs. More specifically, the SLG/FLG flakes can be used as friction-free "ion slides" for organic electrolytes, namely 1 M TEABF$_4$ in propylene carbonate (PC). Based on this concept, we extended these studies for the case of an electrolyte based on organic solvent/ionic electrolyte mixtures (see details in Supporting Information, Section 8). As shown by the tribological measurement reported in Figure S7, the SLG/FLG flakes reduce the nanoscale roughness of the *sprayed*-AC electrodes, decreasing the electrolyte friction by squeezing out the interfacial solid-like region formed by the confinement of the electrolyte within micropores and nanorough surfaces [38,127–131].

### 2.3. Electrochemical characterisation

In the previous sections, we have shown that both the multimodal pore size distribution and the chemical, electrical and tribological properties of the active materials could represent synergistic key-factors to increase the electrochemically accessible area of the electrodes [132]. In order to demonstrate this concept and provide practical guidelines to design advanced EDLCs, the electrochemical performance of the sprayed and casted electrodes, precisely *sprayed*-AC, *sprayed*-AC:SLG/FL, *casted*-AC and *casted*-AC:SLG/FL, were initially evaluated in symmetric coin cell-type EDLCs (see details in Supporting Information) by cyclic voltammetry (CV), galvanostatic charge/discharge (GCD) and electrochemical impedance spectroscopy (EIS) measurements. γ-Butyrolactone/[EMIM][TFSI] was used as electrolyte capable of operating with electrochemical stability within large voltage windows (similar to ionic liquids) [133,134] over a wide temperature range (−40/+100 °C) [135,136]. Figure S8 reports the data obtained for *sprayed*-AC:SLG/FLG adopting different active material weight ratios (*i.e.*, 90:10 and 50:50 wt/wt), showing that the optimal performances were obtained for the electrode with a weight ratio of 90:10 (whose measurements are therefore hereafter shown). Fig. 3a,b show the CV curves of the EDLCs at the scan rates of 0.1 and 5 V s$^{-1}$, respectively. The voltage range of 3.0 V was selected as the upper limit operative voltage ($V^+$) of γ-Butyrolactone/[EMIM][TFSI] to prevent parasitic redox reactions due to the electrolyte degradation [135,136]. At 0.1 V s$^{-1}$ (low scan rate) (Fig. 3a), the quasi-rectangular shape of CV curves without any redox peaks confirms the double-layer capacitive behaviour of the electrodes in the corresponding voltage ranges [3,137]. The non-appearance of current peaks/shoulders associated to faradaic reactions suggests the absence of (reduced) graphene oxides ((R)GO), which typically shows pseudo-capacitive behaviours [138–141]. Noteworthy, although the pseudo-capacitive behaviour of RGO can boost the specific energy in SCs, it could limit both the rate capability and cyclic life of the devices [142–144], consequently impeding high-power applications (as instead targeted by EDLCs). Moreover, the oxygen functionalities of RGO, as well as its structural disorder originated by the reduction process, have been demonstrated to degrade the ultra-low friction/super-lubricating properties of the pristine graphene (see discussion in the previous section and supporting information) [102,103,145], which instead act as "ion slides" for the realization of EDLCs with an ideal rate-independent high specific energy [38]. To prove the advantage of our SLG/FLG flakes as active material compared to a commercial RGO, sprayed hybrid electrodes with commercial RGO replacing our SLG/FLG flakes were also fabricated and characterized by using identical protocols (Figure S9). At high voltage scan rate, the data show that the RGO-based sprayed hybrid EDLC (*sprayed*-AC:RGO) displays a specific current lower than *sprayed*-AC:SLG/FLG (by ~48% at $V^+/2$ for voltage scan rate of 5 V s$^{-1}$), as a consequence of its different morphology (highly wrinkled flakes) and structural properties (reduced 2D-order of the sp$^2$ lattice compared to the one of the SLG/FLG flakes) (Figure S10). Moreover, the *sprayed*-AC:SLG/FLG shows a specific current (~2.3 A g$^{-1}$ at $V^+/2$, forward scans) significantly higher than those measured for the casted EDLCs (< 2 A g$^{-1}$ at $V^+/2$, forward scan). This indicates that the spray coating deposition of the electrode materials is effective to achieve an optimal electrode morphology with electrochemically accessible surface area higher than that of the typical casted EDLCs. These results are consistent with SSA$_{BET}$ data. At 5 V s$^{-1}$ (high scan rate), see Fig. 3b, the hybridized active materials significantly increase the specific current of the *sprayed*-AC (the same trend is observed in casted EDLCs). Notably, the addition of the SLG/FLG flakes reduces the resistive losses, which are evidenced by the shrinkage of the bi-convex (lens-shaped) CV curves. Such shape is caused by the resistive voltage losses ($\Delta V_{losses}$) originating from the electrical resistance of the active material films and current collectors ($R_s$) [137,146], and the electrolyte resistance across the nano/mesoporous structure of the electrodes ($R_{el}$) [137,146–149]. Notably, according to Ohm's law (*i.e.*, $\Delta V_{losses} = (R_s + R_{el}) \times i$), these effects are observable in EDLCs adopting commercial-like mass loadings (of the order of few mg cm$^{-2}$), since they result in high currents at high voltage scan rates (> 0.5 V s$^{-1}$) (*i.e.*, "areal" current densities higher than tens/hundreds of mA cm$^{-2}$ at specific powers typically targeted by EDLCs). According to these considerations, the CV data unambiguously indicate that the SLG/FLG flakes reduce the $\Delta V_{losses}$ of the *sprayed*-AC as a consequence of the decrease of the electrical resistivity of the electrodes (lowering $R_s$) and the favourable tribological properties, *i.e.*, friction-free-like ion movement (lowering $R_{el}$). The effects of the SLG/FLG flakes are even more evidenced by GCD curves acquired at 10 A g$^{-1}$ (Fig. 3c). In fact, the resistive voltage drop (*i.e.*, $\Delta V_{losses}$) at the half-cycle increases up to values comparable to $V^+$ for *sprayed*-AC, which consequently starts to exhibit a resistor-like behaviour. In contrast, *sprayed*-AC:SLG/FLG still shows a capacitive behaviour. In addition, at 0.1 A g$^{-1}$ (*i.e.*, low specific current) (see Figure S11), the *sprayed*-AC:SLG/FLG shows a discharge time ($t_d$) of ~756 s, which is significantly higher (+46.2%) than the one of the *casted*-AC:SLG/FLG (~517 s). In agreement with the Equation (3), the $t_d$ trend indicates a superior electrode $C_g$ of the *sprayed*-AC:SLG/FLG compared to the *casted*-AC:SLG/FLG (~101 F g$^{-1}$ vs. 69 F g$^{-1}$), as expected from CV (Fig. 3a) and SSA$_{BET}$ analyses. Figure S12 reports the electrode $C_g$ and volumetric capacitance ($C_v$) of the EDLCs calculated from their GCD curves at specific currents between 0.1 and 20 A g$^{-1}$. These data further confirm how the spray coating deposition of the electrode materials maximizes the electrode $C_g$ and $C_v$, while the incorporation of the SLG/FLG flakes into the active material improves the rate capability of the EDLCs. While the maximum applicable specific current for the *sprayed*-AC is 20 A g$^{-1}$, the electrode $C_g$ of the *sprayed*-AC:SLG/FLG is still measurable at 50 A g$^{-1}$ (4.7 F g$^{-1}$). Fig. 3d,b show the Ragone plots (in terms of gravimetric and volumetric quantities, respectively) for both the sprayed and the casted EDLCs, made of from both AC and AC:SLG/FLG as active materials. In particular, at specific powers above 15 kW kg$^{-1}$ and power density above 6 kW L$^{-1}$, the *sprayed*-AC:SLG/FLG exhibits specific energy/energy density higher than that of *sprayed*-AC (*e.g.*, specific energy +275% at a specific power of 30 kW kg$^{-1}$). A similar SLG/FLG flakes-enabled enhancement of the electrochemical performance is observed in the casted EDLCs, as previously demonstrated in conventional EDLC electrolyte (*i.e.*, 1 M TEABF$_4$ in PC) [38]. However, the casted EDLCs show inferior electrochemical performance (*i.e.*, specific energy < 20 Wh kg$^{-1}$ at specific power > 4 kW kg$^{-1}$; energy density < 4 Wh L$^{-1}$ at power density > 7 kW L$^{-1}$) compared with the sprayed EDLCs. Lastly, the *sprayed*-AC:SLG/FLG also displays a rate capability significantly superior to the one achieved by *sprayed*-AC:RGO (see Figure S9c-e).

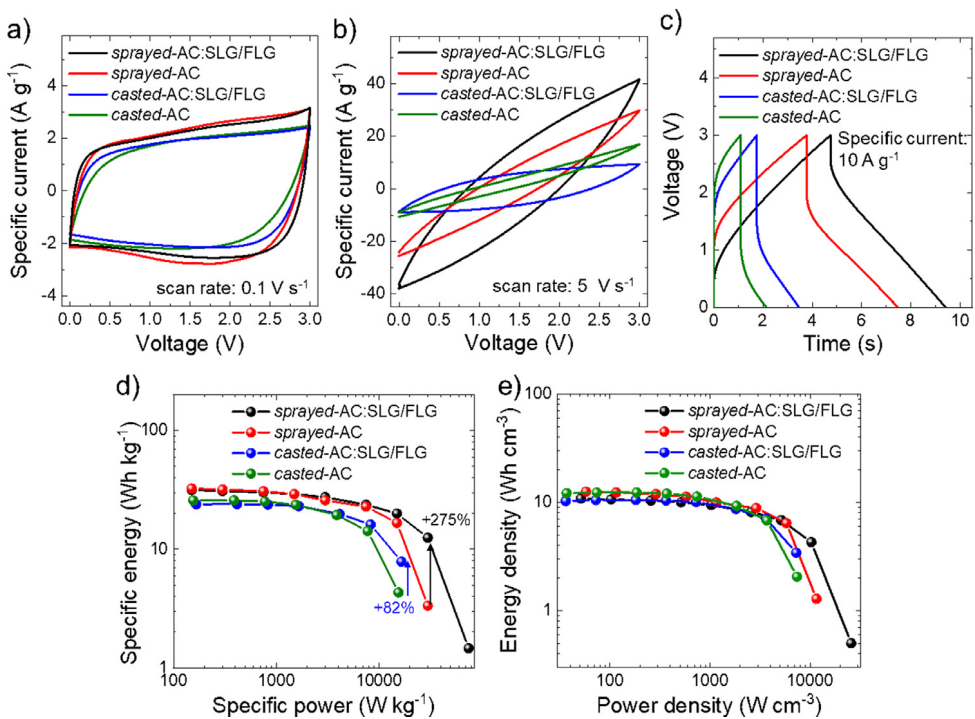

**Fig. 3.** a,b) CV curves of the sprayed and casted EDLCs at voltage scan rates of 0.1 and 5 V s$^{-1}$, respectively. c) GCD curves of the sprayed and casted EDLCs at a specific current of 10 A g$^{-1}$. d,e) Comparison between the Ragone plots of the sprayed and casted EDLCs. Panel d) reports the specific (gravimetric) energy vs. specific power plot, while panel e) displays the (volumetric) energy density *vs.* power density.

The superior electrochemical performance of the *sprayed*-AC:SLG/FLG indicates that the long-range 2D-order of the sp$^2$ lattice of the SLG/FLG flakes (see Figure S3), not observed in RGO (see Figure S10) [150–152], has a dual functional role: i) to increase the electrical conductivity of the electrode material film, as supported by R$_{sheet}$ measurements (see Section 3.2) and ii) to speed up the ion transport within the mesoporous EDLC electrodes, as suggested by LFM analysis (see Figure S7). The EIS analysis of the sprayed EDLCs (Figure S13) further supports the functional roles of the SLF/FLG flakes. The interpretations of EIS data are typically based on physical intuitions and the use of equivalent RC circuits. However, the resulting interpretations are not unique and have often been inconsistent in the literature [153], indicating the need of cautions for the EIS data analysis [154]. As detailed in Supporting Information by referring to relevant literature [153–164], our data support that the SLF/FLG flakes in *sprayed*-AC:SLG/FLG improve the electrical contact between the active material film and the current collector, and decrease the whole internal resistance (R$_{int}$) compared to the case of *sprayed*-AC. In conclusion, the above presented electrochemical measurements highlight that the distinctive morphological, chemical and structural properties of the SLG/FLG flakes can be used to design SLG/FLG-based EDLCs outperforming the EDLCs based on solely AC or hybrid AC/RGO as active materials. Meanwhile, the spray coating deposition of the electrode materials significantly improves the EDLC performances (specific energy and power) of commercial-like SCs produced by slurry-coating methods because of the superior SSA shown by the sprayed electrode films.

*2.4. Validation of the EDLCs over −40/+100 °C temperature range*

To widen the application range of EDLCs, it is necessary to develop a system that withstands harsh environmental conditions, such as both low- and high-temperatures. Typically, commercial EDLCs based on organic electrolytes are specified to work in the –40 to +65/70 °C temperatures range [165,166], although advanced commercial EDLCs can now reach temperature up to 85 °C [165–167]. However, a control of temperature below +70 °C is always recommended to extend the EDLC lifetime, while temperatures inferior to –20 °C can significantly decrease the EDLC electrochemical performance [165,166]. Ionic liquids have been proposed as EDLC electrolyte operating at temperature higher than +100 °C [168,169], however they show minimum operating temperature significantly superior to those of the organic electrolytes. In this context, $\gamma$-Butyrolactone/[EMIM][TFSI] mixture has been successfully proposed to stably operate in a wide range of temperature from –40 to +100 °C [135]. However, the effects of the active materials on the temperature dependence of the EDLC performance was not investigated. In the previous sections, we have shown that the tribological properties of the active material play a significant role in determining the rate capability of the EDLCs. In particular, we have shown that SLG/FLG flakes act as lubricants for the electrolyte, accelerating ions transport through microporous and mesoporous electrode films. To investigate the effect of the SLG/FLG on the EDLC performance at temperature compatible with the $\gamma$-Butyrolactone/[EMIM][TFSI], CV and GCD analyses of our sprayed EDLCs (*i.e.*, *sprayed*-AC and *sprayed*-AC:SLG/FLG) were carried out by controlling the temperature between –40 and +100 °C. The operating voltage window of our EDLCs for the investigated temperatures was assessed by CV analysis. In fact, the temperature affects the electrochemical stability of electrolytic media, typically decreasing their decomposition voltage with increasing the temperature [135,170]. Meanwhile, the electrode materials can also influence the electrochemical stability of the electrolyte. Fig. 4a shows the CV data for *sprayed*-AC:SLG/FLG measured at different temperatures with a low voltage scan rate of 100 mV s$^{-1}$, at which possible faradaic reactions can be detected by whisker-like features when reaching the voltage decomposing the electrolyte [171,172]. On the basis of these CV data, the working potential windows of 4, 3.5 were used as safe conditions to guarantee the electrochemical stability of the EDLCs at –40, 0 °C, respectively, while a potential window of 3 V was used for the EDLCs operating at +25 (room temperature), +50 and +100 °C. Noteworthy, the CV curves exhibit rectangular-like shapes at temperatures ≥ 0 °C, while the lowest tested temperature of –40 °C leads to a lens-shaped CV curve, as a consequence of the high ionic resistance due to the viscosity increase of the electrolyte (approaching to 70 mPa s) [135]. As expected by the increase of the electrolyte viscosity with decreasing the temperature,

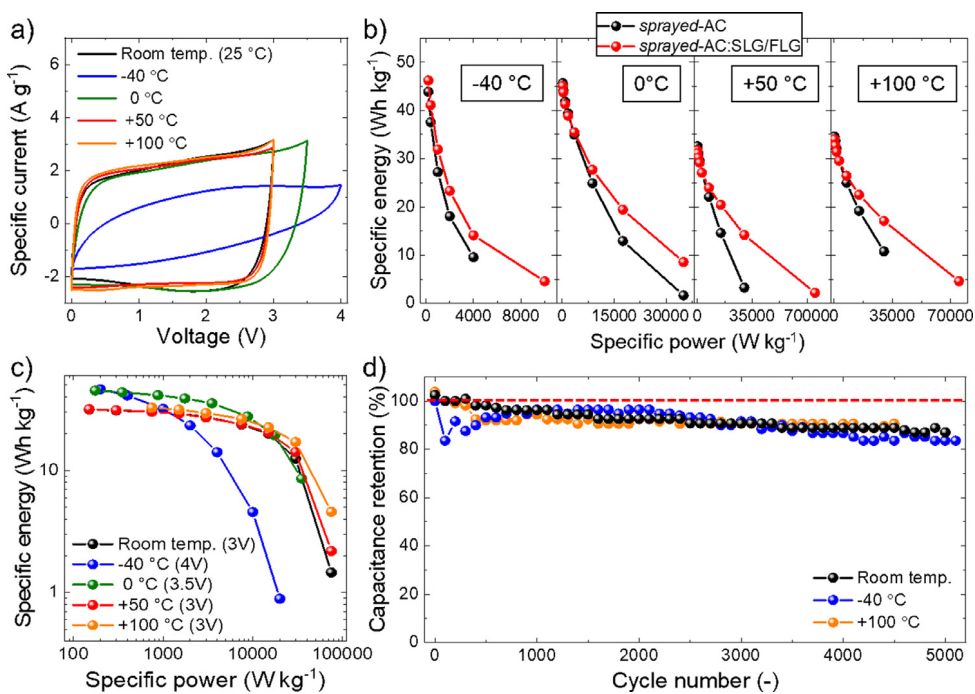

**Fig. 4.** a) CV curves of the *sprayed*-AC:SLG/FLG at a voltage scan rate of 0.1 V s$^{-1}$ and temperature of −40, 0, 25, 50 and 100 °C. b) Electrode $C_g$ of the *sprayed*-AC and *sprayed*-AC:SLG/FLG as a function of the specific current, as calculated by the analysis of their GCD curves. c) Ragone plot of the *sprayed*-AC:SLG/FLG at different temperatures, *i.e.*, −40, 0, 25, 50 and 100 °C. d) Cyclic stability of the *sprayed*-AC:SLG/FLG at a specific current of 1 A g$^{-1}$ at −40, 25 and 100 °C.

Figure S14 shows that the electrode $C_g$ (as calculated by the GCD curve analysis) monotonically decreases with decreasing the temperature. This effect becomes more pronounced as the specific current increases. Interestingly, the SLG/FLG flakes have a beneficial role for increasing the specific current at which the EDLCs start to show a more resistive behaviour. In particular, at −40 °C, the *sprayed*-AC:SLG/FLG still shows an electrode $C_g$ of 8.2 F g$^{-1}$, at 5 A g$^{-1}$, while, as a comparison, the *sprayed*-AC behaves as a resistor for specific currents higher than 2 A g$^{-1}$. The superior performance of *sprayed*-AC:SLG/FLG compared to *sprayed*-AC are further evidenced by comparing their Ragone plots at the investigated temperatures (Fig. 4b). Such enhancement in the EDLC performance in the presence of the SLG/FLG flakes is reasonably attributed to their distinctive tribological properties, making them "slides" for ions, in agreement with the LFM analysis (see Section 3.3), as well as to their high electrical conductivity, decreasing the electrical resistance of the electrode films (see $R_{sheet}$ analysis).

It is worth noticing that, although the electrolyte viscosity at low temperature negatively affects the capacitance of the EDLCs, the observed increase of the electrochemical stability voltage window with decreasing temperature (see Fig. 4a) maximises the specific energy (see Equation (6) in Supporting Information). This can be seen in Fig. 4c that reports the Ragone plot obtained for the *sprayed*-AC:SLG/FLG at different temperatures. It shows that the EDLC delivers the highest specific energy at 0 and −40 °C (*e.g.*, 45.6 and 43.9 Wh kg$^{-1}$ at specific power of 200 W kg$^{-1}$). However, by increasing the specific power above 2 kW kg$^{-1}$, a significant drop of the specific energy is observed for the EDLC operating at −40 °C. On the contrary, the temperature of 0 °C is an optimal working temperature for the cell up to the specific power of 10 kW kg$^{-1}$. Above this specific power (and up to 75 kW kg$^{-1}$), the EDLC working at +100 °C displays the highest specific energy (~4.6 Wh kg$^{-1}$). The Ragone plots obtained for *sprayed*-AC at various temperatures are reported in Figure S15, showing inferior electrochemical performances than those measured for *sprayed*-AC:SLG/FLG, in agreement with the significant role of the SLG/FLG flakes at increasing the EDLC rate capability in a wide range temperature demonstrated in Fig. 4b. Finally, Fig. 4d reports the cyclic stability of the *sprayed*-AC:SLG/FLG at a specific current of 1 A g$^{-1}$ at room temperature, as well as at the most severe conditions, *i.e.*, −40 and +100 °C. The devices retain 87%, 83% and 90% of the initial capacitance over 5000 GCD cycles at room temperature, −40 °C and +100 °C, respectively (values calculated without considering any device pre-treatment). Prospectively, the reduction of the operating voltage window, *i.e.*, the limiting of the $V^+$, could further improve the current stability performance.

### 2.5. Scale-up of the sprayed-AC:SLG/FLF EDLCs

The scale-up of novel EDLCs technologies is mandatory for their practical success, culminating in advanced commercial products. In our case, the WJM exfoliation of graphite enables the production of SLG/FLG flakes up to 0.56 kg per day. This value refers to a single lab-scale WJM apparatus, and prospectively could be scaled up by using multiple WJM machines. Our formulation of the electrode material inks is also promptly scalable since it does not involve any complicated and time-consuming step (see detail in the Supporting Information). To date, more than 10 L of the optimised inks (Figure S16) can be easily produced in a few hours by the sole use of laboratory-scale facilities. Lastly, the EDLCs electrodes have to be fabricated through high-throughput methods, which have to reduce at minimum the human health/environmental impact. Being forced by legislative drivers [80], our electrode material inks have been formulated in "green" solvent, *i.e.*, EtOH/H$_2$O (50:50 vol/vol) mixture, thus enabling their prompt test in industrial pilot-line. Furthermore, our solution naturally eliminates the costs related to the use and recycling of expensive organic solvents (cost of NMP >2 US$ kg$^{-1}$) and related binders (cost of PDVF ranges between 8 and 10 US$ kg$^{-1}$) [82]. Although the NMP is still used during the WJM process to produce the SLG/FLG flakes, it is worth to point out that more than 90% of this NMP is recycled during the subsequent steps needed to extract the SLG/FLG powder. Moreover, we have recently demonstrated that WJM exfoliation of graphite can also be carried out in solvents such as propylene carbonate and propylene glycol, which are cheaper (< 1.5 US$ kg$^{-1}$) and more environmentally friendly than NMP [173]. To definitively demonstrate the feasible industrialisation of our EDLC technology, the spray coating deposition of our optimised inks was carried out by using an automated spray coater, as developed by Thales and M-SOLV to meet high-volume and large-area manufacturing requirements (see technical details in Supporting Information). Importantly, the spray coating set-up is being implemented in high-throughput R2R configuration, which is more appropriate for

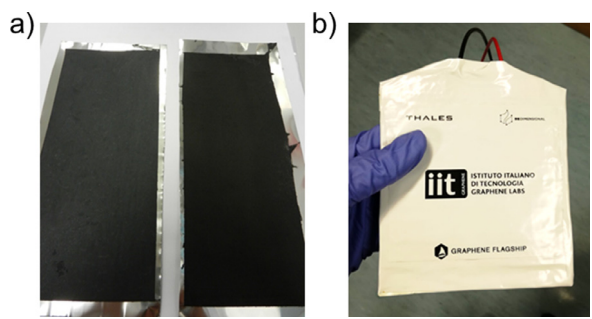

**Fig. 5.** Photographs of a) the 27 cm × 9 cm sprayed electrodes and b) an EDLC stack (three 10 cm × 10 cm EDLCs connected in parallel and providing an overall EDLC stack capacitance of 75 F).

an industrial-scale EDLC manufacturing. Fig. 5a shows the picture of a large-area (27 cm × 9 cm) sprayed electrodes, which macroscopically show homogeneous electrode material deposition. By connecting in parallel three symmetric pouch-cell type EDLCs (obtained by three folding of two 9 cm × 27 cm electrodes using a PVDF membrane as separator), 10 cm × 10 cm EDLC stacks were assembled as prototypes (products exposed to Mobile World Congress 2019 held in Barcelona) (Fig. 5b) [174]. The electrode $C_g$ of a single EDLC is ~85.8 F g$^{-1}$ (absolute values of ~100 F), corresponding to a capacitance of the whole EDLC stack of 75 F (calculated at a scan rate of 10 mV s$^{-1}$). Importantly, the electrode $C_g$ of these large-area electrodes resembles the one measured for lab-scale *sprayed*–AC:SLG/FLG in coin cell-type configurations. These results unambiguously demonstrate that our EDLC technology is promptly scalable, making it ready for its implementation in EDLC pilot-line production.

## 3. Conclusions

We have reported the formulation of a highly scalable and sprayable "green" electrode material-based ink based on activated carbon (AC) and single-/few-layer graphene (SLG/FGL) as an active material for electrochemical double-layer capacitors (EDLCs). By exfoliating graphite through wet-jet milling (WJM) process, SLG/FLG flakes can be produced with an industrial-like rate of more than 0.5 kg per day on a single WJM apparatus. The as-produced inks have been deposited by both manual and automated spray coating in the form of EDLC electrodes. This approach is effective for depositing highly homogeneous electrode films with mass loadings required by current EDLC industry for practical applications (between 1 and 10 mg cm$^{-2}$). By studying the morphological/structural/chemical properties of our electrodes, we have unravelled that spray coating deposition of the as-formulated inks maximises the electrolyte accessibility to the electrode surface area, while SLG/FLG may act as "ion slides" for EDLC electrolytes, reducing the friction of ions over electrode films. Therefore, our sprayed graphene-based EDLCs display superior rate capability (specific energy > 31.5 Wh kg$^{-1}$ at a specific power of 150 W kg$^{-1}$ and > 12.5 Wh kg$^{-1}$ at 30 kW kg$^{-1}$) compared to both graphene-free devices (*e.g.*, specific energy < 4 Wh kg$^{-1}$ at specific power of 30 kW kg$^{-1}$) and commercial-like SCs produced by slurry-coating methods (specific energy < 20 Wh kg$^{-1}$ at specific power > 4 kW kg$^{-1}$). The replacement of SLG/FLG flakes with commercial reduced graphene oxide in the ink composition causes a deterioration of the rate capability attained by the sprayed EDLCs, indicating the defect-free structure and flat morphology of the SLG/FLG flakes are crucial to provide suitable electrical and tribological properties accelerating the ion transport compared to AC films and guarantying high electrochemically accessible SSA at high specific power operation. The distinctive properties of pristine graphene, as well as the spray-coating-based electrode fabrication, allow the EDLCs to efficiently operate in a wide range of temperatures (−40/+100 °C) compatible with ionic liquid/organic solvent mixture-based electrolytes, overcoming the power density limits of graphene-free EDLCs. A prototype EDLCs stack consisting of multiple large-area EDLCs, each one displaying a capacitance of 25 F, is realized to demonstrate the readiness of the proposed technology for an industrial implementation. These findings can give insights into the development of scalable and environmentally friendly EDLC materials/production methods [175] alternative and outperforming those adopted by current EDLC industry.

## Declaration of Competing Interest

The authors declare no conflict of interest.

## CRediT authorship contribution statement

**Mohammad Akbari Garakani:** Conceptualization, Investigation, Writing - original draft. **Sebastiano Bellani:** Conceptualization, Investigation, Writing - original draft. **Vittorio Pellegrini:** Supervision, Writing - review & editing, Resources. **Reinier Oropesa-Nuñez:** Investigation, Methodology. **Antonio Esau Del Rio Castillo:** Investigation, Methodology. **Sara Abouali:** Investigation, Data curation, Formal analysis. **Leyla Najafi:** Data curation, Formal analysis. **Beatriz Martín-García:** Data curation, Formal analysis. **Alberto Ansaldo:** Data curation, Formal analysis. **Paolo Bondavalli:** Data curation, Formal analysis. **Cansunur Demirci:** Data curation, Formal analysis. **Valentino Romano:** Data curation, Formal analysis. **Elisa Mantero:** Investigation, Methodology. **Luigi Marasco:** Investigation, Methodology. **Mirko Prato:** Investigation, Methodology. **Gaetan Bracciale:** Investigation, Methodology. **Francesco Bonaccorso:** Supervision, Writing - review & editing, Resources.

## Acknowledgements

This project has received funding from the European Union's Horizon 2020 research and innovation program under grant agreements No.785219-GrapheneCore2 and No. 881603-GrapheneCore3. This project has received funding from European Union's MSCA-ITN ULTIMATE project under grant agreement No. 813036. We thank M-SOLV for the realization of the spray-coating machine used in this work and for the continuous technical support. We thank the Materials Characterization Facility, Istituto Italiano di Tecnologia, for support in AFM/XPS measurements; the Electron Microscopy facility – Istituto Italiano di Tecnologia for support in TEM data acquisition; and IIT Clean Room facility – Istituto Italiano di Tecnologia for the access to carry out SEM characterisation.

## Supplementary materials

Supplementary material associated with this article can be found, in the online version, at doi:10.1016/j.ensm.2020.08.036.